\documentclass[aps,twocolumn,showpacs,showkeys,floatfix]{revtex4}

\usepackage{graphicx}
\usepackage{dcolumn}
\usepackage{bm}

\newcommand{\Tr}{\mbox{Tr}}
\newcommand{\grad}{\mbox{grad}}
\renewcommand{\div}{\mbox{div}}
\newcommand{\pdiffl}[2]{\frac{\partial #1}{\partial #2}}
\newcommand{\ldiffl}[2]{\frac{D #1}{D #2}}
\newcommand{\diffl}[2]{\frac{d #1}{d #2}}

\begin{document}


\title{Crystal Plasticity for Dynamic Loading at High Pressures and Strains}

\date{December 29, 2007, revised March 17, 2008}

\author{Damian C. Swift}
\email{dswift@llnl.gov}
\affiliation{%
   Material Science and Technology Division,
   Lawrence Livermore National Laboratory,
   7000 East Avenue, Livermore, CA~94551, USA
}

\author{Eric N. Loomis}
\affiliation{%
   Group P-24, Physics Division, Los Alamos National Laboratory,
   MS~E526, Los Alamos, New Mexico 87545, USA
}

\author{Pedro Peralta}
\affiliation{%
   Department of Mechanical and Aerospace Engineering, Arizona State University,
   P.O. Box 876106, Tempe, Arizona 85287, USA
}

\author{Bassem El-Dasher}
\affiliation{%
   Material~Science and Technology Division,
   Lawrence Livermore National Laboratory,
   7000 East Avenue, Livermore, CA~94551, USA
}

\begin{abstract}
A crystal plasticity theory was developed for use in simulations of
dynamic loading at high pressures and strain rates.
At pressures of the order of the bulk modulus, compressions $o(100\%)$
may be induced.
At strain rates $o(10^9)$/s or higher, elastic strains may reach $o(10\%)$,
which may change the orientation of the slip systems significantly
with respect to the stress field.
Elastic strain rather than stress was used in defining the local state,
providing a more direct connection with electronic structure predictions
and consistency with the treatment of compression in initial value problems in
continuum dynamics.
Plastic flow was treated through explicit slip systems,
with flow on each system taken to occur by thermally-activated random jumps
biased by the resolved stress.
Compared with simple Arrhenius rates,
the biased random jumps caused significant differences in plastic strain rate
as a function of temperature and pressure, and provided a seamless
transition to the ultimate theoretical strength of the material.
The behavior of the theory was investigated for matter with approximate
properties for Ta,
demonstrating the importance of the high pressure,
high strain rate contributions.
\end{abstract}


\maketitle

\section{Introduction}
Plasticity has been investigated extensively in terms of slip systems for
deformation at relatively slow rates, close to ambient pressure 
\cite{Rashid92}.
Some treatments incorporate thermal activation of dislocation glide,
but are usually still constructed around an empirical yield stress
for each slip system, and an empirical (power-law) hardening model
\cite{Schoenfeld98,Clayton05,Zhang05,Vogler08}.
In contrast, deformation at strain rates appropriate for shock loading
is almost exclusively described in terms of isotropic elasticity and  
macroscopic, empirical flow theories,
usually treating plastic strain through a single, scalar parameter
\cite{Steinberg-Guinan,Steinberg-Lund,MTS1,MTS2,PTW}.
These formulations are sub-Hookean, treating pressure and stress in a 
fundamentally different way which is an unnatural framework to extend to
anisotropic elasticity.
Plastic flow theories based explicitly on slip systems have only
recently been developed in a form suitable for simulating dynamic loading
to moderate pressure (tens of percent of the bulk modulus),
again using a flow rule for each system based on a yield stress
\cite{Becker04}.

In principle, it is possible to predict any material properties given
an adequately detailed description of the composition and microstructure.
At present, there are few properties that can be predicted as accurately
as they can be measured by experiment: this is a deficiency in our
understanding of condensed matter physics.
Specifically, it is not yet possible to predict the properties of
polycrystalline `engineering' materials to the accuracy where structures
and devices can be constructed reliably and safely, without experiments
to determine the material properties.
One important application is the seeding of instabilities in inertial
confinement fusion (ICF) implosions by spatial variations in the response of the
thermonuclear fuel capsule \cite{Swift_pop}.
In current ICF designs, plastic flow occurs during loading to several
hundred gigapascals over time scales of a few nanoseconds: conditions that are
far more extreme than considered in other applications involving plastic flow.
Measurements of the velocity history and {\it in-situ} x-ray diffraction
during shock loading of single crystals of Be on nanosecond time scales
showed features more consistent with a pure thermal activation model
rather than models based on a yield stress \cite{Swift_Be_07},
so the present study focuses on the construction of a crystal plasticity
theory suitable for simulations of extreme conditions of pressure and
strain rate using thermal activation rather than yield stresses.

More generally, laser ablation is used increasingly in studies of the 
response of materials to dynamic loading, often exploring pressures 
to hundreds of gigapascals with durations of order 1\,ns or shorter,
and investigating the effect of crystal orientation and microstructure.
Such experiments can only be interpreted in the context of high-pressure,
high-rate crystal plasticity, and care is needed to consider the physics
of deformation in adequate detail particularly when drawing conclusions or
applying models on different (generally longer) time scales.

It is reasonable to suppose that a quantitative understanding of
polycrystalline materials can be achieved through an understanding of
the individual crystals making up the microstructure.
The focus of the research reported here is the development of a theoretical
approach to plasticity allowing the response of material microstructures
to be predicted given relevant properties of the single crystals,
for deformation at high pressures and on short time scales.
Materials with a grain size of order micrometers or less can be
simulated for short periods using molecular dynamics (MD) \cite{Bringa}, 
but this is impractical for the huge class of materials with larger grain sizes
or where the response must be followed for long periods.
The interatomic potentials used in MD simulations are not always accurate,
particularly for non-close-packed crystal structures.
Dislocation dynamics (DD) simulations can treat larger volumes of material,
but again the interaction and evolution rules for the dislocations are 
not necessarily accurate.
The complementary approach adopted here is to simulate the response of 
polycrystalline 
materials using continuum dynamics, where the microstructure is resolved
and treated in terms of the response of individual crystals.
As well as investigating physical processes occurring in dynamic deformation
on length scales from single crystals to polycrystalline microstructures,
this type of model will help in the identification of key material parameters
and the development of physics-based models with reduced sets of parameters
for continuum simulations in which the microstructure is not resolved. 
Resolved-microstructure simulations have been reported, and compared with
experimental observations \cite{Schoenfeld98,Clayton05,Zhang05}, 
but not at strain rates and peak pressures high enough to apply to
ICF.

Because the conditions of interest are so far removed from the focus of
previous research in crystal plasticity, it is important to preserve as much
as possible of the simpler material models and numerical methods previously
applied to such conditions;
this constraint guided the structure of the crystal plasticity theory
described here.
We chose to retain the deviatoric decomposition of
stress and strain used for isotropic strength models \cite{Benson92}, 
allowing the isotropic pressure-compression relation to be captured by a
scalar equation of state (EOS), which is well-researched and reliable
compared with the non-isotropic elastic and plastic response 
at high pressures and strain rates.
Similarly, we followed an operator-splitting algorithm \cite{Benson92} for the
simultaneous integration of elastic and plastic strain rates,
widely used in the shock regime and suitable for stiff systems of equations.

Pressures of the order of the bulk modulus induce compressions of tens of
percent, which generally cause significant changes in elastic constants
and plastic flow rates.
Modern dynamic loading experiments commonly explore material response on 
nanosecond to picosecond time scales, where elastic strains may reach
of order 10\%\ \cite{Swift APL};
under these conditions the orientation of slip systems
may change by several degrees even with no rotation of the material,
so resolved shear stresses may change significantly.
Here we present a theory of plastic flow derived consistently for 
high pressures and large elastic strains.
Calculations of the importance of various contributions to the behavior
of matter in these extreme conditions are demonstrated for Ta,
which has been investigated experimentally and using other theoretical
approaches, with many of the basic properties such as elastic constants
predicted at high pressures.
The purpose of this paper is not to present a complete and accurate
theory for plasticity in Ta in the regime of interest,
but to investigate novel properties of plastic flow under extreme conditions.

\section{Continuum mechanics}
The problem of principal interest is to predict the response of
spatially-resolved samples of known properties under the influence of
a load which may vary with time and position.
At each location in the sample at each instant of time, the material
is described by its velocity $\vec u$ and state $s$.
This is an initial value problem: given $\{\vec u,s\}(\vec r,t_0)$ 
over some region $\{\vec r\}\in {\cal R}^3$, 
what is $\{\vec u,s\}(\vec r,t > t_0)$?
Dynamic loading may be induced by the impact of material in different
regions of the problem, or by boundary conditions such as constrained
values of $\vec u(\vec r,t)$ or the stress tensor $\tau(\vec r,t)$.
for $\{\vec r\} \in d{\cal R}^3$.
(Note that, throughout this work,
the stress is defined in the local rest frame of the material,
i.e. it is the Cauchy stress.)
In condensed matter on sub-microsecond time scales, heat conduction
is often too slow to have a significant effect on the response of the material,
and is ignored here.
The equations of non-relativistic continuum dynamics describe the conservation 
of mass, momentum, and energy for matter fields with negligible diffusion
of atoms, evolving under external and internal stresses.
In Lagrangian form, i.e. along characteristics moving with the local material
velocity $\vec u(\vec r)$,
\begin{eqnarray}
\ldiffl{\rho(\vec r,t)}t & = & -\rho(\vec r,t)\div\,\vec u(\vec r,t) \\
\ldiffl{\vec u(\vec r,t)}t & = & \frac 1{\rho(\vec r,t)}\div\,\tau(\vec r,t)\\
\ldiffl{i(\vec r,t)}t & = & \frac 1{\rho(\vec r,t)}
   ||\tau(\vec r,t)\grad\,\vec u(\vec r,t)||
\end{eqnarray}
\cite{Benson92}
where $\rho$ is the mass density and $i$ the specific internal energy.

We follow the frame-based convention common in compressible hydrodynamics that
`Lagrangian' refers to a frame moving locally with the material and
`Eulerian' refers to a frame fixed in space.
This is in contrast to the derivative-based convention common in 
continuum mechanics that
`Lagrangean' refers to derivatives with respect to the original, undeformed
coordinate system and ~Eulerian' refers either to derivatives with respect to
coordinates fixed in space or to coordinates moving locally with the material.
Our meaning of `Lagrangian' derivatives is with respect to coordinates
translated along with the material, but not compressed or sheared.
This convention leads to choices of tensor definitions which are somewhat
different from those usually seen in continuum mechanics,
but make it significantly easier to perform integrations with respect to time 
over general large-deformation paths
relevant to loading at high pressures and strain rates.

The continuum dynamics equations are closed through the addition of
constitutive relations describing the inherent properties of each material:
$\{\tau(s);\dot s(s,\grad\,\vec u,\partial i/\partial t)\}$ 
where $s(\vec r,t)$ is the field describing the local state of the material.
The second law of thermodynamics provides a constraint
on the constitutive model.
$\tau(s)$ is the EOS; $\dot s(s,\grad\,\vec u,\partial i/\partial t)$ describes
the evolution of the state under an applied deformation and heating rate.
As well as experiencing compression and work from mechanical deformation,
$s(\vec r,t)$ can evolve through internal processes such as plastic flow.
Subsuming $D\rho/Dt$ and $Di/Dt$ into the evolution relation,
\begin{equation}
\ldiffl{s(\vec r,t)}t \equiv \dot s[s(\vec r,t),U(\vec r,t)]
   \quad:\quad U\equiv \grad\,\vec u(\vec r,t).
\end{equation}
The crystal plasticity model described here was implemented in a
software package of material models, 
providing a common interface for different varieties of
continuum dynamics program \cite{Ariadne,Swift_genscalar_07}.
The software structure is object-oriented, the continuum dynamics program 
storing the local material state $s$ regardless of type, 
and the material package performing calculations including $\tau(s)$
and the rate of change of state given $\grad\,\vec u$ or 
$\partial i/\partial t$.

There is an inconsistency in the standard continuum dynamics treatment
of scalar (pressure) and tensor (stress) response.
The scalar EOS expresses the pressure $p(\rho,T)$ as the 
dependent quantity, which is the most convenient form for use in the
continuum equations.
Standard practice is to use sub-Hookean elasticity \cite{Benson92}, in which the
stress tensor is found by integration
\begin{equation}
\dot\tau = c\dot e
\end{equation}
where $c$ are elastic constants 
and $\dot e$ is the elastic strain rate tensor.
Thus the isotropic and deviatoric contributions to stress are not
treated in an equivalent way: the pressure is calculated from a local state 
involving a strain-like parameter (mass density), whereas the deviatoric stress
is part of the local state and evolves by time-integration.
This is an undesirable distinction which complicates the problem.
The stress can be considered as a direct response of the
material to the instantaneous state of elastic strain: $\tau(e,T)$.
This relation can be predicted directly with electronic structure calculations
of the stress tensor in a solid for a given compression and elastic strain
state \cite{Poirier99}, 
and is a direct generalization of the scalar EOS.

The constitutive response of a crystal is described most naturally in terms
of a coordinate system fixed with respect to the symmetry directions of
the crystal, such as its lattice vectors.
The orientation of elements of material varies with position and time in
the problem, and is described completely by a local rotation between
crystal and problem coordinates.
Thus the state $s$ is chosen to give the elastic strain $e$ in the coordinate
frame of the crystal, 
and includes the rotation matrix $R$ from the crystal to the
problem coordinates.
This is a polar decomposition of the instantaneous deformation gradient 
$F=R^TV$,
where $e$ is related to the right stretch tensor $V$ through a specific choice
of finite strain measure.
The evolution equations for $V$ (hence $e$) and $R$ implicitly include
contributions from the `external' evolution of the continuum
and from the `internal'
evolution of the local material state by plastic flow: subscripts $c$ and $p$
respectively:
\begin{equation}
\dot V = \dot V_c + \dot V_p,\quad
\dot e = \dot e_c + \dot e_p,\quad \dot R = \dot R_c + \dot R_p.
\end{equation}
Integrating over time, this is a multiplicative decomposition of the
deformation gradient $F$ into elastic and plastic contributions, though
only the elastic contribution is required explicitly.

At any instant of time, the continuum dynamics calculation provides
$\grad\,\vec u(\vec r,t)\equiv U$.
This is decomposed into symmetric and antisymmetric terms $S$ and $A$,
respectively describing strain and rotation rates:
\begin{equation}
U = S + A \quad : \quad S = \frac 12(U+U^T), A = \frac 12(U-U^T).
\end{equation}
Neglecting plastic flow for the moment,
the rate of change of the right stretch tensor $V$ from the evolution of
the continuum is
\begin{equation}
\dot V = RSR^TV.
\end{equation}
The external rotation rate is found similarly by noting that 
the incremental rotation caused by $A$ over a short increment of time is
$\delta R_c\simeq I+\delta t A^T$ to order $\delta t$, so
\begin{equation}
\dot R_c = A^TR+RA.
\end{equation}

The strain tensor $e$ is calculated from $V$, or $\dot e_c$ from $\dot V_c$,
through the use of a finite strain measure.
The main consideration in the choice of strain measure is consistency with
the measure chosen when inferring the stress tensor $\tau$ from measured
or calculated states of finite strain.
Possible choices include $\ln V$ and $(V^m-I)/m$, where $m=2$ gives
the Green-St~Venant measure \cite{Ogden}.
The strain is objective, which makes
the formulation computationally convenient.

The formulation used for the constitutive model was that,
for a given composition and crystal structure of a material,
in the frame of reference of the crystal,
which is rotated with respect to the continuum dynamics frame,
the stress tensor $\tau_x$ depends only on the elastic strain tensor $e$
and the temperature $T$.
Strictly, $\tau_x$ also depends on the defect structure in the material;
for most situations of practical or research interest, the vast majority of
atoms in a crystal are in an environment which has the underlying symmetry
of the crystal, so the direct contribution of defects to $\tau_x$ was ignored.
Plastic flow acts to reduce shear stresses, i.e. to reduce the shear
components of elastic strain.
Plastic flow is mediated by the motion of defects such as dislocations,
defined with respect to the instantaneous crystal axes.
At the level of the crystal lattice, the plastic state comprises the population
of defects $\{\phi\}$ on each slip system
and the population of obstacles to defect motion $\{\omega\}$.
The right stretch tensor is needed in order to relate the defect lattice
vectors to the instantaneous elastic strain state.
For the strain measures considered, it is more efficient to
calculate $e(V)$ than $V(e)$, so $V$ was used in defining the material state.
Thus the local, instantaneous state of the material is
$s(\vec r,t)=\{R,V,T,\{\phi\},\{\omega\}\}$.
The constitutive relation describes $\tau_x(s)$,
the relaxation of elastic strain by plastic flow $\dot e_p(s)$
and the resulting conversion of elastic to thermal energy $\dot T(s,\dot e_p)$,
and the evolution of the plastic state as flow occurs:
$\{\dot\phi\}(s,\dot e_p)$ and $\{\dot\omega\}(s,\dot e_p)$.
A hyperelastic stress-strain relation was used:
\begin{equation}
\tau_x = c(s)e
\end{equation}
where $c(s)$ is the tensor of elastic `constants,' 
explicit functions of elastic strain (particularly compression) and temperature.
It is easier to ensure that the stress is a path-independent function of
elastic strain with a hyperelastic relation than a hypoelastic relation
\begin{equation}
\dot\tau_x = \tilde c(s)\dot e.
\end{equation}
The (objective) stress in the frame of the problem is found using the
local rotation:
\begin{equation}
\tau(\vec r,t)=R^T(\vec r,t)\tau_x(\vec r,t)R(\vec r,t).
\end{equation}

The effect of plastic flow can be calculated and integrated simultaneously
with the flow equations through $\dot s(s,\grad\,\vec u,\partial i/\partial t)$.
For time-integration by finite difference numerical schemes,
the natural time increment from the continuum dynamics may be very different
from that required for plastic flow.
Similarly, plastic heating may change the state enough that numerical
schemes integrating forward in time become unstable.
For this reason, for most purposes
plastic flow was operator-split from the continuum dynamics \cite{Benson92}:
the continuum equations were integrated at constant plastic state,
then plastic flow was calculated at constant total strain.
For a complicated material response in which the rate of change of
generalized state is a sum of logically-distinct
contributions 
\begin{equation}
\dot s =\sum_i\dot s_i,
\end{equation}
operator-splitting is the use of the approximate integral relation
\begin{equation}
\int_t^{t+\Delta t}\dot s(s)\,dt\simeq
\sum_i\int_t^{t+\Delta t}\dot s_i(\tilde s_i)\,dt
\end{equation}
where $\tilde s_i$ is the result of the previous integrations $j < i$,
i.e. a sequence of integrations over the full time increment $\Delta t$
in which all but one of the contributions in turn are ignored.
This decomposition is used widely in continuum dynamics involving shocks and
other high-rate flows, for example to decouple Eulerian remaps in different
dimensions from each other and from Lagrangian integrations \cite{Benson92}, 
and to decouple stiff chemistry from the continuum in reactive flow 
\cite{OranBoris}.
It is typically second-order accurate in time.
A previously-reported algorithm employed a plastic predictor and an elastic
corrector \cite{Fotiu96} and was demonstrated to be stable and accurate.
However, this is not a natural framework for plastic flow as a response to
dynamic loading at extremely high rates and compressions,
as the material response is best regarded as elastic with plasticity occurring 
as a stress-relaxing mechanism, as demonstrated by observations such
as the decay of elastic precursor waves in shock loading
\cite{Jones62,Taylor63,Stepanov85,Resseguier98}.
Our formulation also appears to be more suitable for deviatoric strength
with an EOS in the strong shock regime.
Operator-splitting allows the plasticity algorithm to be more stable 
than is generally possible when integrated simultaneously with 
explicit-time hydrodynamics (because plastic flow cannot then lead to
time increments that are inconsistent between the continuum and the local
plastic state) with little change in accuracy and similar or better 
computational efficiency.
When integrating the continuum equations with operator-splitting,
the constitutive model was split into a series of partial integrations
\begin{eqnarray}
s_e & = & s(t)
    + \int_{t;s(t)}^{t'} \dot s_e(s,\grad\,\vec u,\partial i_e/\partial t)\,dt \\
s_h & = & s_e
    + \int_{t;s_e}^{t'} \dot s_h(s,{\bf 0},\partial i_h/\partial t)\,dt \\
s_p & = & s_h(t')
    + \int_{t;s_h}^{t'} \dot s_p(s,{\bf 0},\partial i_p/\partial t)\,dt,
\end{eqnarray}
where $\Delta t=t'-t$ is the finite time increment,
$\dot s_e$ is the state evolution under the influence of the
continuum velocity gradient with no plastic flow and heating $\partial i_e/\partial t$ 
restricted to isentropic compression plus kinematic viscous forces,
$\dot s_h$ is heating $\partial i_h/\partial t$ from external sources 
(including radiation deposition,
conduction, and artificial viscosity used to stabilize shock waves),
and $\dot s_p$ is plastic relaxation with the total strain held constant
with heating $\partial i_p/\partial t$ only from dissipation of relieved elastic strain energy.
Integration proceeds by taking $s(t')=s_p$.

The software implementation made wide use of object-oriented techniques,
particularly polymorphism and inheritance.
This allows the constituents of the material models to be chosen when
the program is run, in particular the form and type of the EOS,
and the type of general functions such as polynomials or tabulations.
The ability to define functions in a general way in the input to a program,
rather than having to be specific in the source code, is convenient for
representing the evolution of defect populations in the crystal
(including dislocations and pinning sites) as it is less clear which
precise functional forms to use.

\section{Deviatoric elasticity}
In high pressure loading, plastic flow is usually fast enough that the
isotropic pressure has greater magnitude than shear components of stress.
As discussed above, unusually large elastic shear strains may reach
$\sim$10\%, and are more commonly $\sim$1\%.
At pressures the order of 100\,GPa, isotropic compressions are typically
several tens of percent, so the isotropic stress clearly dominates.
For this reason, we use the deviatoric formulation of elasticity.
The stress and strain tensors $\tau$ and $e$ are decomposed into isotropic and
deviatoric contributions:
\begin{eqnarray}
e(\vec r,t)&\equiv&\epsilon(\vec r,t)+\mu(\vec r,t)I 
\quad:\quad\mu = \frac 13\Tr\,e \\
\tau(s)&\equiv&\sigma(s)-p(s)I\quad:\quad p = -\frac 13\Tr\,\tau.
\end{eqnarray}
The mean pressure $p(s)$ is represented by a scalar EOS,
in practice using the mass density $\rho(\vec r,t)$ rather than
the compression $\mu(\vec r,t)$ in the state $s(\vec r,t)$.
The deviatoric stress-strain relation is
\begin{equation}
\sigma(s)=C(s)\epsilon
\end{equation}
\cite{Hill}
where $C$ is the tensor of deviatoric elastic `constants,'
explicit functions of $\rho$, $T$, and in principle $\epsilon$.
$C$ are obtained from $c$ by subtracting the isentropic bulk modulus $B$ 
from all elements affecting the normal components of stress.
Here we take
\begin{equation}
B\equiv \left.\pdiffl p\rho\right|_s,
\end{equation}
which is well-defined for all crystal systems.
(In general for non-cubic systems, $\epsilon\ne 0$ for $\sigma=0$.)
Voigt notation was used for strain and stress tensors, so $c$ and $C$
were $6\times 6$ matrices.
In Voigt notation,
\begin{equation}
C=c-\left(
\begin{array}{cccccc}
B & B & B & 0 & 0 & 0 \\
B & B & B & 0 & 0 & 0 \\
B & B & B & 0 & 0 & 0 \\
0 & 0 & 0 & 0 & 0 & 0 \\
0 & 0 & 0 & 0 & 0 & 0 \\
0 & 0 & 0 & 0 & 0 & 0 \\
\end{array}
\right).
\end{equation}
Non-linear elasticity is accommodated by
allowing $C$ or $c$ to be a function of the elastic strain deviator
-- this is potentially important for elastic strains above the percent level.

Strictly, the scalar EOS is also a function of the
strain deviator \cite{Swift APL}.
Ignoring this dependence may lead to $o(1\%)$ errors at the elastic strains
considered here.
This refinement will be considered further in future work.

Thermodynamic consistency depends on the structure of the constitutive model.
In thermoelasticity, the entropy at any strain is a function of
temperature only.
Once the defect structure associated with plasticity is included,
it also contributes to the entropy.  However, defects generally 
affect only a small fraction of the atoms.
Therefore, above the Debye temperature of the material,
the defect entropy is small compared with the lattice-vibrational entropy.
In this work, we neglected the entropy of the defects, 
and used thermodynamically consistent scalar EOS to capture the
relation between specific internal energy $i$ and temperature $T$.
For consistency with the construction of the scalar EOS, $i$ was chosen
to exclude the elastic distortional energy, which was stored and conserved
implicitly through the elastic strain.

The scalar EOS has been obtained for many materials
\cite{Kinslow70,SESAME,Steinberg96},
by high-pressure experiments such as impact-induced shock waves
\cite{Eremets96},
and from electronic structure calculations \cite{Swift PRB 01}.
Elastic constants have been measured at STP for many materials,
and can also be predicted as a function of compression,
temperature, and strain from electronic structure calculations
\cite{Poirier99}.
We took one such prediction for Ta up to 1\,TPa \cite{Moriarty02}.
The reported variation of $(C_{11}-C_{12})/2$ with pressure was used to
calculate $c_{11}(\rho)$ and $c_{12}(\rho)$,
using a semi-empirical scalar EOS \cite{SESAME}
to obtain pressure $p(\rho)$ and bulk sound speed $c^2(\rho)$ 
along the zero-kelvin isotherm.
The elastic constants predicted at zero pressure deviated by around 10\%\ from
experimental measurements \cite{Featherston63}.

Conventionally, elastic constants $C$ are expressed as functions of
pressure $p$ and temperature $T$ \cite{Steinberg-Guinan,Becker04}.
In normal materials, $\partial C/\partial p|_T>0$
and $\partial C/\partial T|_p<0$, with 
$|\partial C/\partial p|_T|>|\partial C/\partial T|_p|$.
It is more convenient to calculate $C(\rho,T)$ from electronic structure
calculations rather than $C(p,T)$, and generally
$|\partial C/\partial T|_\rho|<|\partial C/\partial T|_p|$.
In fact, in some cases 
it may be reasonable to ignore the temperature-dependence of
$C(\rho,T)$.
For instance, considering the variation of the polycrystalline shear
modulus $G$ for Ta \cite{Steinberg96}, 
$\partial G/\partial p|_T=1.45 10^{-2} G_0$/GPa and
$\partial G/\partial T|_p=-1.3 10^{-4} G_0$/K,
implying
$\partial G/\partial T|_\rho=-1.298 10^{-4} G_0$/K.
The decomposition in terms of $\rho$ and $T$ is therefore at least as 
appropriate for elasticity over a wide range of compressions.

\section{Crystal plasticity}
At the level of a single crystal or grain, 
plastic flow is mediated by defects -- dislocations or disclinations --
which act along a set of slip systems defined with respect to the lattice
vectors.
The theory described here is based on previous work on rate-dependent
crystal plasticity \cite{Rashid92}, 
but generalized to treat high compressions and elastic strains,
which entails the inclusion of some corrections and refinements.
The theory was developed for dislocation-mediated plasticity,
but is expressed in a form which enables it to be used to represent
flow by twinning, in at least a simplified manner.
Where appropriate, we refer to defects rather than dislocations, to make
the point that much of the theory should apply to disclinations.

In general, twinning and slip are potentially competing mechanisms which
interact with each other.
The orientation of dislocations and slip systems changes when material twins.
The presence of dislocations can impede the motion of a twin boundary,
the structure and mobility of a dislocation are altered if it passes through
a twin boundary, and the presence of twinned material alters the strain 
field from, and forces on, a nearby dislocation.
The treatment of plasticity adopted here is to resolve each element of material
with a single crystal orientation, populated by defects described by a set of
homogeneous densities, mediating strain by their motion.
When applied to twinning, this approach lends itself to simulations where either
the evolution of twins is resolved spatially, or strain occurs only in a single
plane, or the volume fraction of twinned material is small.
These limitations could be removed by allowing volume fractions of each
variant of twin orientation, as has been reported previously 
\cite{Kalidindi98,Staroselsky98,Karaman00,Salem05,Barton07}.
For practical reasons of rapid application to real problems, 
previous models of twinning have been constructed around a yield stress for
each deformation mechanism, with modifications to power-law hardening,
rather than explicit defect kinetics. Simplifications have been made
such as assuming that dislocation motion would be negligible in twinned
regions.
Generally, models of twinning are less advanced than those of crystal 
plasticity, and we present the dislocation-focused work here as 
a starting point for future developments in twinning rather than a complete
framework for a model.

The local state in the material is described by a set of defect populations 
and a set of obstacles to defect motion.
Each defect population is characterized by its type (e.g. edge, screw),
orientation with respect to the crystal lattice vectors 
(e.g. by its Burgers vector $\vec b$), and local density $\phi$.
In reality, extended defects (dislocations and disclinations) are curved,
and may be of mixed type.  In this work, mixed-type or
mixed-orientation defects are accounted for by assigning them proportionately
to the appropriate `pure' populations.
The local state of the material thus comprises the sets of defect 
and obstacle densities, $\{\phi_b\}$ and $\{\omega\}$.
For each population of defects, there is a set of planes and directions 
$\{\vec n,\vec d\}$ 
in which the crystal shears.
Each specific combination of $\vec b$, $\vec n$, and $\vec d$ 
is referred to as a slip
system, but different slip systems share the same defect population if
$\vec b$ is the same.
There are separate defect populations with $\vec b$ and $-\vec b$,
but for given $\vec b$ there are not separate slip systems with
direction $\vec n$ and $-\vec n$ or $\vec d$ and $-\vec d$: 
the latter is treated as reverse motion
on the slip system with direction $\vec d$.
The obstacles experienced by defects with $\vec b$ are the same as those
with $-\vec b$, so
the local state of the material was taken more compactly as
$\{\phi_b,\phi_{\bar b}\};\{\omega\}$.

At the atomic level, the relevant densities of defects and obstacles are
expressed with respect to the densities of atoms.
Thus the defect and obstacle densities do not change with elastic compression.

Under the action of the local stress, the defects move thus causing plastic
flow which relieves elastic strain, and the populations of defects and
obstacles evolve.
The plastic strain rate was calculated for each slip system independently.
The time increment for integrating plastic flow was chosen based on
the total plastic strain rate, so that the increment of plastic strain
was limited to a small value (such as 0.01) per time step.
Since plastic flow was operator-split from the continuum dynamics,
it was thus subcycled at constant total strain over the time step
required by the continuum dynamics.
This approach is used widely for the simultaneous solution of stiff equations
with continuum dynamics, for example in reactive flow \cite{OranBoris}.
In this instance, it greatly simplifies the calculation of potentially
competing slip systems.

\begin{figure*}
\begin{center}
\includegraphics[scale=0.6]{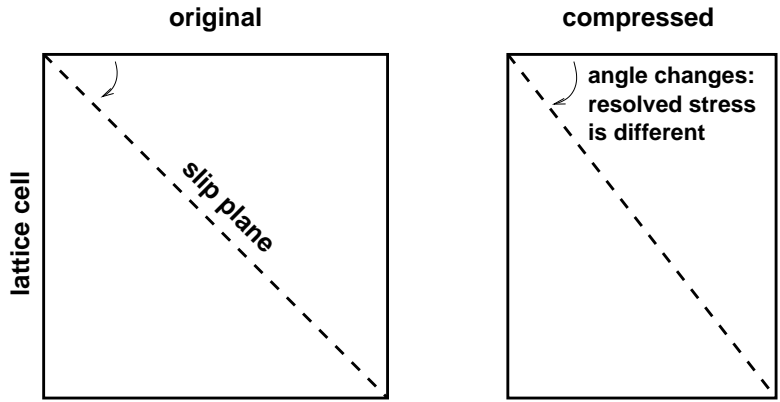}
\end{center}
\caption{Schematic of the effect of finite elastic strain on slip system
   orientations,
   hence on the resolved shear stress and strain rate.}
\label{fig:finitestrain_slipsysts}
\end{figure*}

Given the stress tensor $\tau$, the stress resolved along the slip system is 
\begin{equation}
\tau_r = \vec d\cdot\tau\cdot\vec n = \vec d\cdot\sigma\cdot\vec n = \sigma_r,
\end{equation}
the resolved deviatoric stress,
since $\vec n$ and $\vec d$ are always orthogonal.
In the frame of the undistorted crystal,
slip in a given system produces a change of total strain of
\begin{equation}
\Delta e_T \propto -\vec d\otimes\vec n,
\end{equation}
which is traceless -- no volume change -- 
since $\vec n$ and $\vec d$ are orthogonal.
The strain increment can be split into elastic (unloading) 
and rotational parts,
\begin{eqnarray}
\Delta e_p & = & \Delta \epsilon_p
    \propto -\frac 12\left(\Delta e_T+\Delta e_T^T\right), \\
\Delta R_p & \propto & +\frac 12\left(\Delta e_T-\Delta e_T^T\right),
\end{eqnarray}
where the sign of $\Delta R_p$ is opposite to that of $\Delta e_p$
because $R$ rotates from the crystal coordinates to the instantaneous
orientation in the laboratory space.
The convention chosen here is to calculate
the strain increment with respect to the undistorted lattice,
and accumulate specific or per-atom defect and obstacle densities.
If $\vec n_0$ and $\vec d_0'$ are the normal and direction as real space vectors
with respect to the undeformed crystal, in the deformed crystal
\begin{equation}
\vec d' = V \vec d_0',\quad 
\vec n = \frac{\left(V^{-1}\right)^T \vec n_0}{|\left(V^{-1}\right)^T \vec n_0|},
\end{equation}
where the latter equation uses Nanson's formula.
The change in elastic strain is then calculated with respect
to the distorted lattice, using the right stretch tensor $V$, 
\begin{equation}
\Delta \tilde e_T = d'\otimes\vec n',
\end{equation}
and used to calculate plastic flow increments of stretch $\Delta V_p$
and rotation $\Delta R_p$:
\begin{eqnarray}
\Delta V_p & = & -\frac 12\left(\Delta \tilde e_T+\Delta \tilde e_T^T\right), \\
\Delta R_p & = & +\frac 12\left(\Delta \tilde e_T-\Delta \tilde e_T^T\right),
\end{eqnarray}
This correction takes into account the orientation effect of finite strain
and the scaling of the lattice vectors by isotropic compression
(Fig.~\ref{fig:finitestrain_slipsysts}).

\begin{figure}
\begin{center}
\includegraphics[scale=0.7]{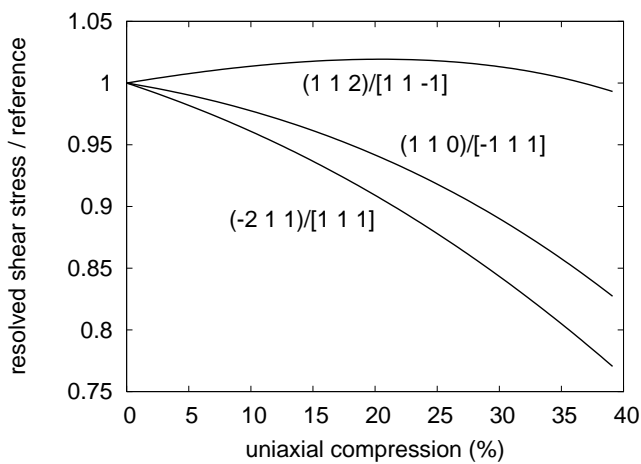}
\end{center}
\caption{Effect of finite elastic strain on shear stress resolved along
   slip systems in Ta, for uniaxial compression along $[100]$.
   Reference value is resolved shear stress on each system neglecting the
   effect of finite strain.}
\label{fig:finitestrain_Ta100}
\end{figure}

In Ta under high-rate loading conditions, screw dislocations dominate the
behavior \cite{Moriarty02}.
There are four distinct dislocation populations (and two signs of each),
with $\vec b=\frac 12\langle 111\rangle$.
Each population with $\pm\vec b$ may slip along 
$\{110\}$ planes (3 systems), $\{112\}$ planes (3 systems), and
$\{123\}$ planes (6 systems)
(see e.g. \cite{NematNasser98} for a full enumeration of the 
Burgers vectors and planes).
For a crystal compressed uniaxially along $[100]$, the effect of finite
elastic strain is an $o(10\%)$ effect for strains of tens of percent
(Fig.~\ref{fig:finitestrain_Ta100}).

Defect motion proceeds as a thermally-activated hop past the Peierls barrier.
In principle, hopping may occur in the direction of increasing shear energy;
it is more likely to occur in the direction of decreasing shear energy.
With no elastic stress, hopping is equally likely in either direction.
An applied stress biases the average hopping direction by altering the
effective Peierls barrier by the work that would be done against the applied
stress at the peak of the barrier.
If the Peierls barrier energy is $E_b$, the effective barrier is
$E_b\pm \Delta E$ where $\Delta E = \sigma_r fl/2$, $l$ is the 
interatomic spacing in the direction of slip (a function of compression),
and $f$ scales $\sigma_r b$ to the units used to express $E_b$, e.g. an
area per atom if $E_b$ is expressed as the energy per atom.
The net hopping rate $\dot\lambda$ should thus follow an Arrhenius form 
with competing forward and reverse contributions,
\begin{equation}
\dot\lambda = Z(\rho,T)\left[
   e^{-\frac{E_b-\Delta E}{k_BT}}-e^{-\frac{E_b+\Delta E}{k_BT}}
   \right]
\end{equation}
where $Z$ is the rate at which the defect attempts to hop --
the vibration frequency of the atoms -- and $k_B$ is Boltzmann's constant.
Rearranging to factorize out the unstrained Arrhenius barrier for
single-direction hopping,
\begin{equation}
\dot\lambda = 2 Z(\rho,T) \exp\left(-\frac{E_b}{k_BT}\right)
   \sinh\frac{\Delta E}{k_BT}.
\end{equation}
Each exponential term is the probability of a successful hop:
the Arrhenius form applies when the probability is less than one,
and each exponential was limited to one.
Strictly, the Arrhenius form applies when the hopping attempts are
described by a Boltzmann distribution.
In a solid, the distribution is given
by the population of phonon modes, which gives a modified probability
particularly below the Debye temperature, when the population of the phonon
modes falls significantly and zero-point energy becomes more evident.
The hyperbolic sine dependence has been discussed previously in the context
of creep \cite{Nichols71}, though the treatment of applied strain and stress
is different here.
When dislocations move, the rate at which a straight dislocation develops
a forward-moving kink is lower than the rate at which the kink spreads
laterally as adjacent sections of the dislocation move forward,
because the energy of an isolated kink is higher.
On average, the net successful hopping rate may depend on hops of the
same direction by several adjacent atoms ($N$ say), so
\begin{equation}
\dot\lambda = Z(\rho,T)\left[
   e^{-N\frac{E_b-\Delta E}{k_BT}}-e^{-N\frac{E_b+\Delta E}{k_BT}}
   \right].
\end{equation}

If there were no obstacles, the overall plastic strain rate on the
slip system would be
\begin{equation}
\dot\alpha_{bnd} = \dot\lambda_{bnd} \left(\phi_b+\phi_{\bar b}\right).
\end{equation}
The effect of obstacles is to reduce the mobility of dislocations
and hence reduce the mean plastic strain rate,
because of the finite energy required to generate
additional lengths of curved dislocations.
This energy includes contributions from the dislocation core and the strain
field, which for a curved dislocation generally includes self-interactions.
Obstacles include point defects (vacancies, substitutional impurities, and
interstitials), other populations of dislocations, and disclinations.
If there were only one type of obstacle (density $\omega$) and it were able
to pin dislocations perfectly,
the resulting strain rate on the slip system would be
\begin{equation}
\dot\alpha_{bnd} = \dot\lambda_{bnd}\left(\phi_b+\phi_{\bar b}\right)P(\omega),
\end{equation}
where $P(0)=1$ and $P(1)=0$.
Here we choose $P(\omega)=1-\omega$ for $0 < \omega < 1$ per atom.
Each type of obstacle has a different inherent effectiveness in reducing
the mobility of dislocations of each type, represented here by a factor
$\beta_{ibnd}$ for the $i$th type of obstacle.
Dislocations may be able to pass obstacles, which can be characterized by
an Arrhenius rate so that the overall effectiveness of a given type of obstacle
is 
\begin{equation}
\tilde\beta_{ibnd} \equiv
   \beta_{ibnd}\left(1-e^{-\frac{E_o-\Delta E}{k_BT}}\right)
\end{equation}
where $E_o$ is the energy barrier for the dislocation to pass the obstacle
and $\Delta E$ is calculated as for Peierls barriers.
Here we assume that the density of obstacles is much lower than that of the
atoms in the crystal, so reverse hopping should be neglected though barrier
lowering is included.
Combining the effects of all obstacles,
\begin{equation}
\dot\alpha_{bnd} = \dot\lambda_{bnd}\left(\phi_b+\phi_{\bar b}\right)
   P(\omega_{bnd}),
\end{equation}
where
\begin{equation}
\omega_{bnd}\equiv\sum_i\tilde\beta_{ibnd}\omega_i.
\end{equation}
Each slip system contributes to the unloading of the elastic strain at a rate
in each system of
\begin{equation}
\dot\epsilon_{bnd} = -\frac 12\dot\alpha_{bnd}
   \left(\vec d\otimes\vec n+\vec n\otimes\vec d\right).
\end{equation}
As plastic flow is expressed as a strain rate which is a function of the
instantaneous, local strain state, this formulation can be regarded as
viscoplastic.

In previous work, the attempt frequency $Z$ for the motion of a dislocation
has been held to be much less than the attempt frequency for an atom
\cite{Moriarty02}.
However, vibrations of the atoms provide the fundamental mechanism for
dislocations to move.
We note that the reverse-hopping contribution and the multiple-atom factor
decrease the net successful hop rate, and could be interpreted as
a smaller attempt frequency in an over-simplified Arrhenius treatment.

\begin{figure}
\begin{center}
\includegraphics[scale=0.7]{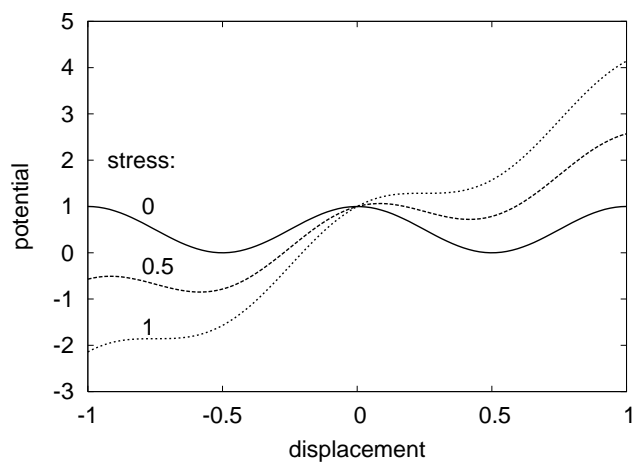}
\end{center}
\caption{Effect of resolved stress on the periodic potential
   experienced by a moving defect (displacement in units of the
   interatomic spacing, amplitude in units of the Peierls barrier,
   stress in units of the ultimate strength).
   An applied stress reduces the barrier height in the direction of
   (mean) plastic strain; at the ultimate strength the barrier 
   disappears in that direction.}
\label{fig:Peierlsbias}
\end{figure}

In this theory, the effect of the applied stress is to tilt the energy surface
seen by atoms constituting the defect as they hop in the slip direction.
The ultimate strength of the material is given by the maximum gradient
in the unstressed energy surface: when the resolved stress reaches this
level, there is no barrier to defect motion.
For a sine wave potential,
\begin{equation}
V(x)=\frac{E_b}2\sin\frac{2\pi x}l,
\end{equation}
the ultimate strength is $\pi E_b/l$.
(Fig.~\ref{fig:Peierlsbias}.)

\begin{figure}
\begin{center}
\includegraphics[scale=0.7]{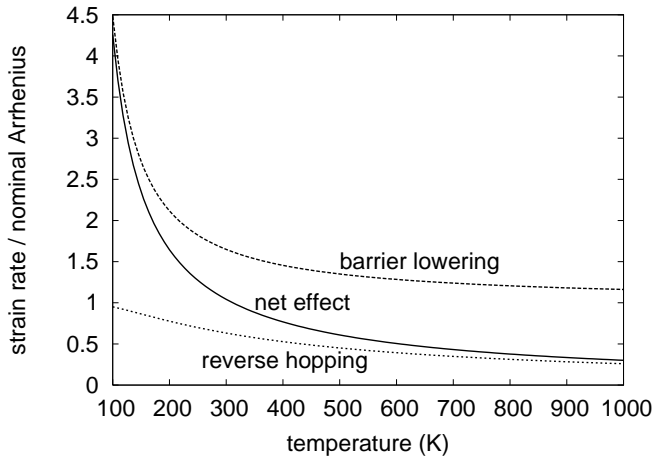}
\end{center}
\caption{Effect of the barrier-lowering and reverse-hopping terms 
   on the strain rate, for Ta at zero pressure with 1\,GPa resolved stress.}
\label{fig:hopcontrib}
\end{figure}

The reverse-hop term has not been included in other plasticity studies
using Arrhenius-based flow rates.
Its effect is greatest at low strains, i.e. at low strain rates,
where the stress biasing term is smallest.
For example, in Ta 
(flow stress observed in microsecond-scale loading experiments
$\sim$1\,GPa \cite{Steinberg96},
distance between adjacent Peierls valleys $l \simeq 1.5 a$ where
$a=3.308$\,\AA), 
the effect of the reverse term is to reduce the strain rate by
tens of percent for temperatures over a few hundred Kelvin,
and the lowering of the barrier by the applied stress 
increases the strain rate by several tens of percent.
Comparing the overall effect with the nominal Arrhenius hop rate,
the net effect is to increase the strain rate by hundreds of percent 
at temperatures below about 310\,K and to decrease it by tens of
percent at higher temperatures.
(Fig.~\ref{fig:hopcontrib}).

\begin{figure}
\begin{center}
\includegraphics[scale=0.7]{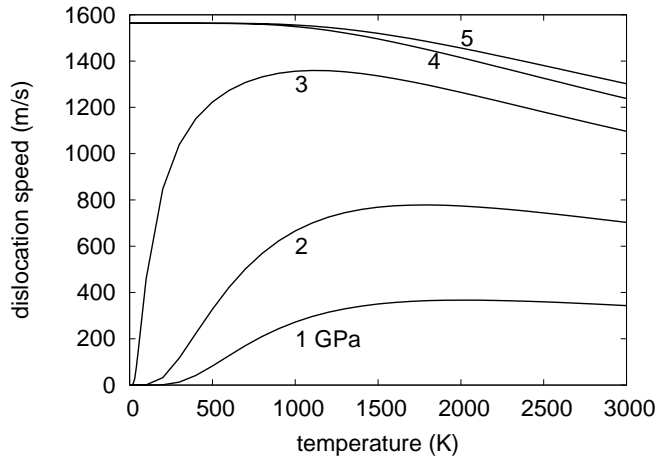}
\end{center}
\caption{Predicted effect of resolved shear stress (GPa) and temperature
   on the average speed of $\langle 111\rangle\{110\}$ dislocation,
   for Ta at 16.7\,g/cm$^2$.}
\label{fig:Ta_dislocspeed1}
\end{figure}

The effect of stress and temperature can also be seen by calculating
the average speed of a dislocation moving through the lattice, 
\begin{equation}
\vec u_d \equiv \dot\lambda |\vec b|
\end{equation}
For temperatures of a few hundred kelvins and resolved shear stress (rss)
well below the Peierls stress, 
the dislocation speed increases with temperature and rss,
and has a value of similar order to kinetic Monte-Carlo predictions
using an explicit representation of the double-kink mechanism
\cite{Deo05}.
However, once the rss exceeds the Peierls stress, the dislocation speed
is limited to a value around half of the bulk sound speed, because of
saturation of the probability for forward jumps.
At high temperatures, the average dislocation speed decreases with
temperature as the probability for reverse jumps increases.
(Fig.~\ref{fig:Ta_dislocspeed1}.)

\begin{figure}
\begin{center}
\includegraphics[scale=0.7]{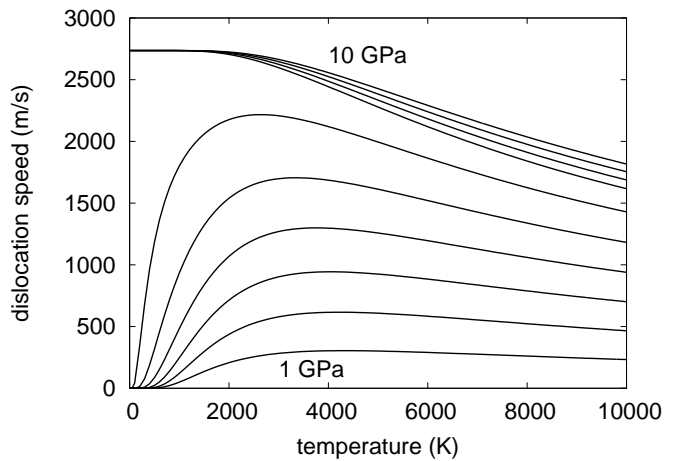}
\end{center}
\caption{Predicted effect of resolved shear stress (GPa) and temperature
   on the average speed of $\langle 111\rangle\{110\}$ dislocation,
   for Ta isentropically compressed to 20\,g/cm$^3$ (around 49\,GPa on the
   principal isentrope).}
\label{fig:Ta_dislocspeed2}
\end{figure}

Repeating these predictions for Ta compressed to 20\,g/cm$^3$,
the same trends are evident, but the saturation stress is greater than
6\,GPa and the maximum dislocation speed is higher in proportion
to the rise in frequency of atomic vibrations (Fig.~\ref{fig:Ta_dislocspeed2}).

The total working rate is $W\equiv||\sigma\dot\epsilon_p||$
where $\epsilon_p \equiv -\sum_{b,n,d}\epsilon_{bnd}$.
This is the rate at which elastic strain energy is relieved;
it is not all converted into heat as some remains stored as the energy of
the defect structures.
The proportion $f_p$ converted to heat is thought to be 85-95\%
\cite{Farren25,Taylor34};
this was treated here as a parameter in the theory,
but could be calculated by subtracting the core and interaction energy of the
defect population, say
\begin{equation}
E\equiv\sum_b\eta_b\phi_b+\sum_{bb'}\upsilon_{bb'}\phi_b\phi_{b'}.
\end{equation}
Such a calculation has been attempted previously, using a yield stress
and power-law hardening model for the crystal, but so far without an accurate
treatment of the scalar EOS or elasticity at high compressions
\cite{Clayton05}.

Since the rate of plastic flow is as a sum over slip systems, and the
rate is calculated independently for each system, this theory automatically
treats non-Schmid behavior where flow may occur on a slip system which does
not have the greatest rss \cite{Qin92}, if it is kinetically favorable.
This condition may occur if the system with greatest rss
has insufficient dislocations to relieve the elastic stress, or if the
Peierls barrier is higher than for a competing system of lower rss.
As the rate of plastic flow on a slip system is non-linear in rss,
when competing slip systems have the same defect populations and kinetics,
flow is dominated by the system with greatest rss, giving normal Schmid
behavior.

As plastic flow occurs, the populations of dislocations and obstacles evolve.
Sections of dislocation may annihilate if they reach a grain boundary
or a dislocation of opposite Burgers vector.
In most situations studied experimentally \cite{Peralta05},
the dominant effect is the creation of additional lengths of dislocation
by Frank-Read sources between obstacles \cite{Hirth}.
Dislocations may be pinned by intersection with other defects in the
lattice such as dislocations in other slip systems.
In the present work,
the dislocation density was taken to evolve according
to the generic relation
\begin{equation}
\begin{array}{rcl}
\dot\phi_b = F_0&+&\left[
      F_1(\phi_b,\omega_b)-F_2(\phi_b \phi_{\bar b})
   \right] \dot\alpha_b \\
   &+&\sum_{b'}R_{bb'}\left(\phi_{b'}\dot\alpha_b+\phi_b\dot\alpha_{b'}\right),
\end{array}
\end{equation}
where 
\begin{equation}
\dot\alpha_b\equiv\sum_{n,d}\dot\alpha_{bnd},
\end{equation}
the $F_i$ are respectively spontaneous dislocation generation,
net dislocation generation by flow
(including annihilation with the grain boundary),
and annihilation with dislocations of opposite Burgers vector,
and the $R_{ij}$ are reaction terms between dislocations of different types,
which may involve Arrhenius rates to cross barriers.
Spontaneous generation of defects is a reversible process limited by
an energy barrier $E_g$
with a rate
\begin{equation}
F_0 = Z(\rho,T) \left[
   e^{-N_g\frac{E_g-\Delta E}{k_BT}}-e^{-N_g\frac{E_g+\Delta E}{k_BT}}
   \right].
\end{equation}
Compared with flow by defect motion, spontaneous generation can affect
any atom in the crystal, but has a high energy barrier.

Dislocation reactions may make a significant difference to the plastic flow
rate.
Mobile dislocations can combine to form sessile dislocations, reducing the
density of mobile dislocations and acting as obstacles to the remaining
mobile dislocations.

Obstacles which are dislocations on other slip systems are described by the
dislocation density on those slip systems.
In this case, the effectiveness parameters $\beta_{ibnd}$ constitute
a hardening matrix.

The vibration rate $Z(\rho,T)$ can be estimated from 
electronic structure calculations
or measured for instance by low-energy neutron scattering.
The Peierls barrier can also be estimated from electronic structure
calculations with appropriate variations in elastic strain,
or inferred from interatomic potentials.
The initial density of defects and obstacles or pins is best obtained by
analysis of the microstructure.
The functions describing the evolution of dislocations and obstacles, 
including the hardening parameters $\beta_{ibnd}$, can be estimated from
MD or DD simulations.
The fraction of plastic work converted to heat could be estimated from
these simulations.

\section{Example simulations of plastic flow in tantalum}
As an illustration of the behavior of constitutive models based on the
theory described above, we consider the time-dependent relaxation of
shear stress in material subjected to an instantaneous uniaxial compression.
This scenario represents the conditions induced by rapid loading such as
laser ablation on a single crystal.
It has been observed experimentally in a variety of materials
including Be \cite{Swift_pop},
and is related to the phenomenon of decay of a supported elastic wave
\cite{Johnson in Asay}.
As before, the material was taken to be `Ta-like,' using parameters taken
from the literature.

The elastic constants and slip systems were as described above.
The frequency factor $Z(\rho,T)$ for defect motion was estimated from
the ambient Debye temperature of Ta.
$Z(\rho)$ was assumed to be dominated by the variation of the elastic
constants, so $Z\sim \sqrt{B/\rho}$, i.e. proportional to the bulk sound speed.
Any temperature variation of $Z$ was neglected.
The Peierls barrier $E_b(\rho,T)$ to defect motion was taken from
the zero-pressure value of 0.18\,eV calculated for rigid shear 
without lattice relaxation
\cite{Moriarty02} and the calculated variation of
the ultimate strength $\tau_c$ of Ta with pressure,
\begin{equation}
\diffl{\tau_c}p = 0.12\diffl Gp
\end{equation}
\cite{Moriarty02}
where $G$ is the shear modulus.
$dG/dp$ was estimated from the pressure-hardening term in the 
Steinberg-Guinan strength model
\cite{Steinberg96},
assuming
\begin{equation}
\frac 1{E_b}\diffl{E_b}p = \frac 1{\tau_c}\diffl{\tau_c}p.
\end{equation}
As with the frequency factor, any explicit temperature-dependence of the
Peierls barrier was neglected.
In both cases, there was an implicit temperature dependence as functions
of $p$, as they were expressed in terms of $\rho$.

The initial dislocation density in any crystalline material depends completely
on the details of preparation: cooling rate from a melt, amount of subsequent 
working, temperature at which working occurred, and so on.
For metal crystals grown from molten material, e.g. by zone melting,
the dislocation density is typically \cite{Elbaum60}
in the range 1 to 10 per $\mu$m$^2$,
which is equivalent to $6$ to $60\times 10^{-8}$ per atom.
The population of each orientation depends on, for example, the temperature
gradient with respect to the lattice vectors 
during the formation of the crystal,
and the orientation of the principal axes of the strain fields applied during
working, again with respect to the lattice vectors.
The dislocation density was taken to be distributed uniformly over all
orientations of dislocation.
Obstacles include dislocations on other slip systems, grain boundaries,
impurities, and interstitials, and also depend completely on the details
of manufacture.
Simulations were performed with different obstacle densities, and different
models of obstacle evolution.
The hardening parameters $\beta_{ibnd}$ followed previous studies
\cite{Rashid92} in taking $\beta=1$ for self-hardening and 
$\beta=1.4$ for cross-hardening.
Dislocation reactions were not considered.

Simulations were performed of Ta crystals compressed along $[100]$ and
$[110]$.
We consider two general classes of simulation, which explore different
aspects of the material response:
an initial elastic strain followed by plastic relaxation
(which closely mirrors the characteristics of shock-loading experiments),
and deformation at a constant strain rate
(which is equivalent to more conventional dynamic loading tests).
Except for the study of initial temperature, the sample was initially at STP.
The orientation was chosen so that compression occurred along the $x$-direction
in the coordinate reference frame.

\subsection{Elastic strain followed by plastic relaxation}
\begin{figure}
\begin{center}
\includegraphics[scale=0.7]{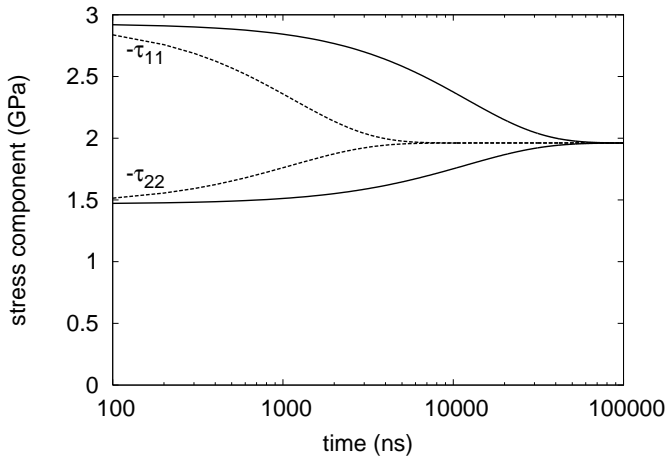}
\end{center}
\caption{Simulated stress relaxation in a Ta crystal compressed elastically
   by 1\%\ along $[100]$,
   with no obstacles and no evolution of the dislocation densities.
   The dislocation density was 10/$\mu$m$^2$ (solid) and 100/$\mu$m$^2$ (dashed),
   distributed uniformly over all systems.}
\label{fig:Ta_uniax100rho}
\end{figure}

Following the initial, instantaneous compression,
the normal and transverse components of stress, 
$\tau_{11}$ and $\tau_{22}=\tau_{33}$ respectively, 
relaxed toward the mean pressure
(Fig.~\ref{fig:Ta_uniax100rho}).
As one would expect from the structure of the model,
increasing the obstacle density decreased the rate of stress relaxation,
and the evolution (increase) of the defect density by interaction
with obstacles increased the rate of stress relaxation.
However, 
the detailed choice of the functional forms and parameters,
which is manifested as work-hardening,
is speculative without guidance from MD and DD simulations,
and will be described separately.

\begin{figure}
\begin{center}
\includegraphics[scale=0.7]{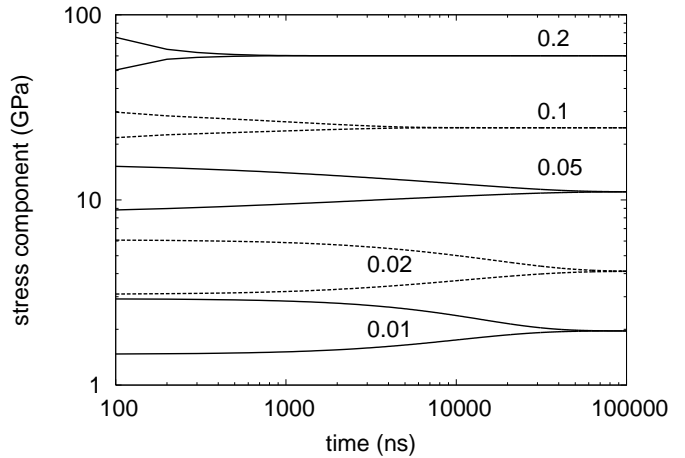}
\end{center}
\caption{Effect of initial compression on simulated stress relaxation
   in a Ta crystal.
   The initial compression was along $[100]$, by fractional amounts
   shown in the graph.
   The dislocation density was held constant at $10/\mu$m$^2$,
   and there were no obstacles.
   Stress components $\tau_{11}$ (upper) and $\tau_{22}$ (lower)
   are shown for each initial compression.}
\label{fig:Ta_uniax100str0}
\end{figure}

\begin{figure}
\begin{center}
\includegraphics[scale=0.7]{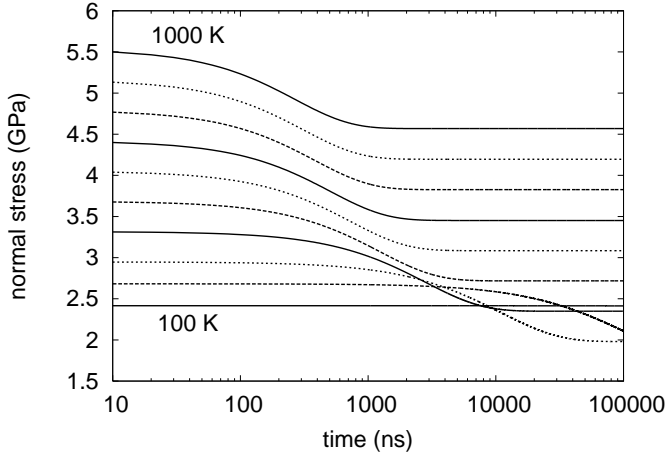}
\end{center}
\caption{Effect of initial temperature (increments of 100\,K)
   on simulated stress relaxation in a Ta crystal.
   Heating was isochoric, so the different initial temperature
   gave different initial pressures.
   The initial compression was 1\%\ along $[100]$.
   The dislocation density was held constant at $10/\mu$m$^2$
   and there were no obstacles.}
\label{fig:Ta_uniax100at}
\end{figure}

Simulations were performed of plastic relaxation for different initial
elastic strains.
For initial strains greater than 5\%, the barrier-lowering effect of the
applied stress gave an abrupt decrease in the time required to relax to
a state of isotropic stress (Fig.~\ref{fig:Ta_uniax100str0}).
Simulations were performed for different values of the initial temperature
from 100 to 1000\,K.
For simplicity, the temperature was applied at the same starting density,
which gave a varying mean pressure.
The time to relax to an isotropic stress state showed a clear
dependence on temperature (Fig.~\ref{fig:Ta_uniax100at}).

\begin{figure}
\begin{center}
\includegraphics[scale=0.7]{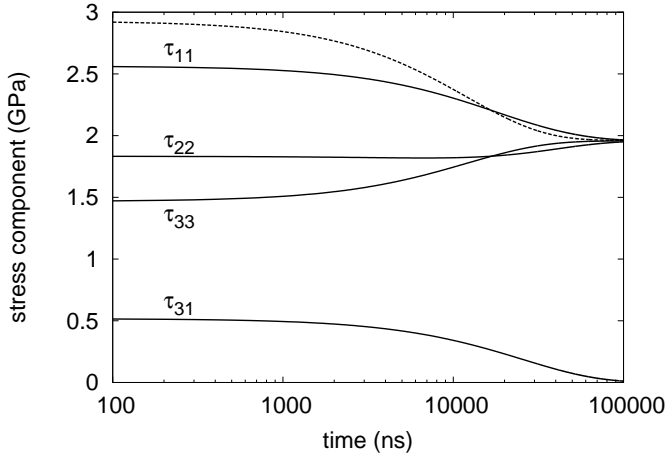}
\end{center}
\caption{Effect of orientation on simulated stress relaxation
   in a Ta crystal.
   The initial compression was 1\%\ along the $x$-axis.
   The dislocation density was held constant at $10/\mu$m$^2$
   and there were no obstacles.
   All non-zero stress components are shown for a crystal with
   $(110)$ planes normal to the $x$-axis;
   the normal stress is shown for a crystal with $(100)$ planes
   normal to the $x$-axis as before (dashed line).}
\label{fig:Ta_uniaxorientcmp}
\end{figure}

The stress relaxation varied significantly between $[100]$ and $[110]$
orientations parallel with the loading direction.
For loading along $[110]$, the initial stress was lower and relaxation
took longer.
There were also significant differences in the other stress components
(Fig.~\ref{fig:Ta_uniaxorientcmp}).

\subsection{Constant strain rate}
\begin{figure}
\begin{center}
\includegraphics[scale=0.7]{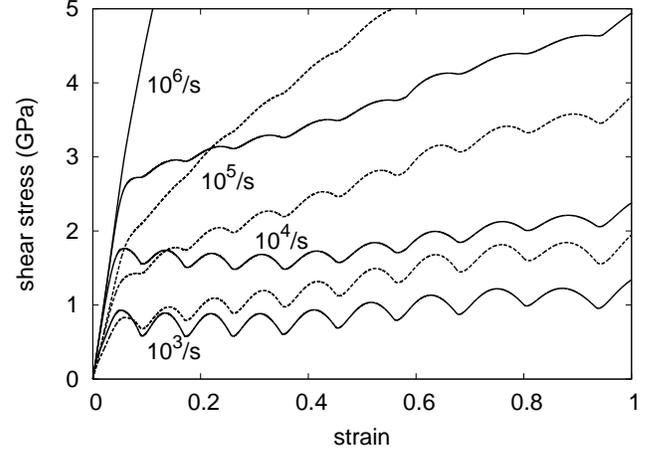}
\end{center}
\caption{Effect of strain rate on shear stress in Ta crystals
   compressed uniaxially along $[100]$ (solid) and $[110]$ (dashed) directions.
   The dislocation density was held constant at $10/\mu$m$^2$
   and there were no obstacles.}
\label{fig:Ta_uniaxratecmp}
\end{figure}

\begin{figure}
\begin{center}
\includegraphics[scale=0.7]{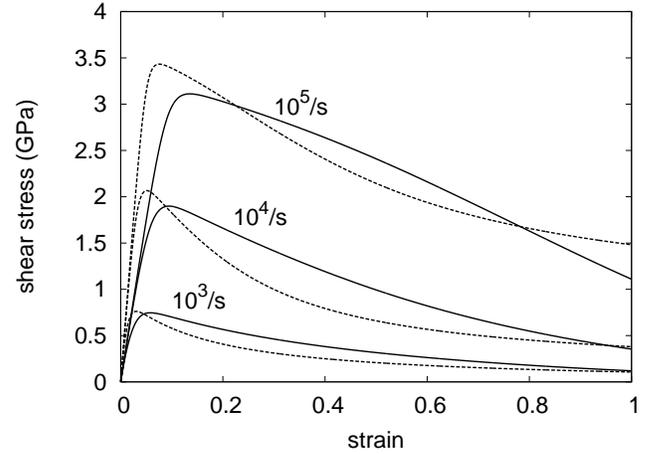}
\end{center}
\caption{Effect of strain rate on shear stress in Ta crystals
   sheared in $[100]$ (solid) and $[110]$ (dashed) planes.
   The dislocation density was held constant at $10/\mu$m$^2$
   and there were no obstacles.}
\label{fig:Ta_shearratecmp}
\end{figure}

During deformation at a constant strain rate,
the finite rate at which defect-mediated plasticity could relax shear stresses
allowed a finite elastic stress to be maintained, i.e. a finite flow stress.
Simulations were performed for deformation at different strain rates,
held constant for a period of time.
Strain rates were applied for uniaxial compression in the $x$-direction 
(representing ramp waves)
and for pure shear in the $xy$ plane.
In uniaxial compression, the shear stress increased with compression
at high strain rates and strains because the elastic constants increased
with compression.
The shear stress exhibited oscillations caused by competition between
thermal softening from plastic heating and pressure hardening from compression.
For shear deformation, the flow stress decreased monotonically with
strain once plastic flow occurred, because of thermal softening.
In both uniaxial and shear cases, there were regions around a few percent
strain where the flow stress dipped because of plastic heating.
A crystal would be susceptible to strain localization under these conditions.
The shear stress increased with strain, but showed a clear dependence on
strain rate and crystal orientation.
The relative magnitude of flow stress in $[100]$ and $[110]$ crystals
changed with strain rate.
(Figs~\ref{fig:Ta_uniaxratecmp} and \ref{fig:Ta_shearratecmp}.)

It is interesting to compare these calculations with DD simulations
and with experimental observations.
One feature of the calculations is that initial flow stresses are in the range
of around 0.7-3.5\,GPa.
The strain rate to achieve a given flow stress depends on the dislocation
density.
DD simulations show plastic yield at a similar stress, but with work-hardening
at larger strains as dislocations tangle \cite{Tang08}.
Projectile impact experiments on cast and machined Ta samples 
millimeters thick
have given low pressure flow stresses between 0.77 and 1.1\,GPa 
\cite{Steinberg96}.
Laser-induced shock experiments on rolled foils of Ta 25-50\,$\mu$m thick
have given flow stresses of around 3\,GPa \cite{Tierney08}.
Thus the calculations presented here, though with a simplified model, 
are quite consistent with experiments on widely ranging time scales 
and microstructures.

\section{Discussion}
The theory and model described here include much of the physics behind
plastic flow.
The purpose of the model is to simulate the response of samples to dynamic loading,
resolving the samples spatially so that each single crystal is represented
typically by tens of numerical cells.
This resolution treats the variation in loading experienced by different
parts of each grain, and allows the variation in response to be
calculated.
However, in the continuum dynamics implementation used so far,
the evolution of the defect population is treated locally in each numerical cell.
In reality, defects such as dislocations may move throughout the grain,
and may also pass into adjacent grains.
In grains with initially few pinning sites, dislocations may travel a long way.
It would be an improvement to treat defect transport in the solution of the
continuum dynamics equations.
Similarly, grain boundaries may act as defect sources.
Source terms are included in the present theory, and could be altered 
in cells in contact with grain boundaries, but transport would be needed to
allow the defects to propagate further than the numerical cell size.

The fraction of plastic work converted to heat, $f_p$,
also known as the Taylor-Quinney factor \cite{Quinney37}, is an interesting and
poorly-understood quantity.
As discussed above, we have outlined a way to estimate $1-f_p$ from the
potential energy of the dislocation population, but we have not yet
investigated the properties of this approach in practice.
The total dislocation core energy can be estimated from the dislocation
density.
It may be possible to quantify the energy associated with pins and other
obstacles in a similar way.
These contributions could then be included directly in the theory.
The elastic strain energy associated with the defects is likely to depend
more sensitively on the configuration of the defects.
It may be possible to estimate it from the dislocation density;
the accuracy could be investigated by comparison with MD
and DD simulations.

The formulation chosen for defect-based flow provides an explicit,
lattice-level mechanism for the Bauschinger effect through the
dependence of the dislocation kinetics on the mass density,
which differs between compression and release,
though this is not the only contribution.
Defect evolution may be such that an initial plastic deformation
reduces the flow stress on reverse loading, but this would depend on the  
relative behavior of the generation and annihilation terms and of the
evolution of obstacle populations.
If rotations are not excessive, it may well be that the initial deformation
could increase the density of some dislocations without increasing the
density of others which would act as obstacles, and then reverse loading
would be accommodated in opposite-direction flow by the increased dislocation
population.
Contributions to the Bauschinger effect may also arise out of deformation of a
multi-grain microstructure.

Although much of the plasticity theory presented here applies to twinning
as well as slip, and although the disclinations separating twinned material
can be viewed as a plane of dislocations, twinning is not fully represented
by this theory in its current form.
At the grain level, twinning occurs by the conversion of material of one
orientation to the twinned orientation.
This re-orientation of the crystal axes means that the slip systems
change in real space, so any competing flow by slip should take this into
account.
If twinning is the dominant mode of deformation, 
and each grain is loaded such that only a single disclination mode is active
in each, then re-orientation does not matter to the material response and
the theory as presented should be able to treat the problem.

We preferred to express the variation of elastic
constants, the Peierls barrier, etc, in terms of the mass density rather than
the pressure.
These quantities can be estimated from electronic structure calculations.
Electronic structure calculations using the local density approximation
for the exchange-correlation energy, which is almost universal,
give results which typically have a systematic error 
$o(1\%)$ in mass density;
this error should be corrected e.g. using the observed mass density at STP 
\cite{Swift PRB 01}.
A particular attraction in the use of electronic structure calculations is
that properties can be predicted for structural phases which are difficult
to examine experimentally, such as those occurring at high pressures
and temperatures induced in shock experiments.

In most of the example simulations above,
a strain was applied instantaneously and the relaxation of the stress state
was predicted.
In contrast, strength models are conventionally tested by applying a constant 
strain rate and predicting the flow stress.
The plasticity theory presented here is motivated by the need to
analyze and interpret high pressure and strain rate experiments 
on single crystals.
Typical experimental data are the velocity history at the surface of 
a planar crystal loaded with a square pressure history.
For loading pressures slightly in excess of the Hugoniot elastic limit,
the velocity history typically includes an elastic wave followed by a
plastic wave.
In single crystals, the rise time of the elastic wave is usually
too rapid to measure \cite{Swift_pop}, but the subsequent
velocity history shows evidence of the time-dependence of plastic flow.
Thus the strain rate during loading is very high and ill-determined;
it is more appropriate in this instance to consider an instantaneous
loading followed by relaxation at constant compression,
as a preliminary to full continuum dynamics simulations in which
the density may also change after the initial compression.

At the high pressures and short time scales considered here, 
the bias effect of the high resolved stresses on the Peierls barrier
can apparently allow shear stresses to relax much more rapidly.
This effect is equivalent to a pressure-dependent reduction in the 
effective flow stress.
This is potentially important in the control of the seeding of implosion 
instabilities of microstructural origin in inertial confinement fusion:
higher-pressure shocks may induce less spatial variation in material
velocity caused by anisotropy in different grains.
According to the simulations for Ta, it is also desirable to have a high
dislocation density.

The philosophy underlying physics-based models of material behavior is to
have general models, parameters inherent to material of a given composition
(e.g. Ta), together with characterization of the microstructure for
each specific application.
The characterization would include the crystallographic texture, 
and defect and obstacle populations.
The plasticity theory described here does not quite follow this ideal,
as some microstructural information is at present implicit in the
functional forms for the evolution of the defect populations.
MD or DD studies are needed to elucidate the generation of
defects and obstacles.

\section{Conclusions}
We have developed a theory for elasticity and plastic flow intended to
represent the response of crystals to deformation at high pressures
and strain rates.
The theory is needed because at pressures over $O(100\,\mbox{GPa})$ and
on time scales shorter than $O(10\,\mbox{ns})$,
elastic strains can exceed several percent, which changes the orientation
of slip systems significantly with respect to the stress field.
The motivation behind this work is to simulate the response of single crystal
and polycrystalline samples to dynamic loading, by resolving the microstructure
explicitly.
The equations describing the response of the material are constructed to be
convenient for integration forward in time.
In this vein, we have preferred to consider the elastic strain rather than
the stress as part of the local material state; this approach is more consistent
with the conventional treatment of the scalar equation of state in continuum
dynamics at high pressures.

For a given total strain, the stress induced by the material was considered
to originate from the elastic strain of the crystal lattice.
Plastic flow was considered to operate at constant total strain, acting to
relieve the elastic stress.
With this approach, arbitrary sets of slip systems can be treated 
simultaneously, and in a spatially-resolved forward-time integration 
there is no need for explicit constraints to ensure elastic compliance.
If a crystal does not possess enough slip systems to relieve part of the
stress, it remains unrelieved unless a grain elsewhere with a different
orientation can slip so as to relieve the total strain.

We have proposed an underlying Arrhenius-based kinetic term for the
motion of dislocations and other defects which mediate plasticity.
Defects do not `know' which way to move: the applied stress was used as a bias
on the Peierls barriers to motion, and the probability of hops in both
directions (increasing as well as decreasing the elastic strain) was
included.
The barrier-lowering effect of the stress was predicted to increase the
plastic strain rate by hundreds of percent at low temperatures,
and the reverse-hopping contribution was predicted to decrease the net
strain rate by tens of percent at high temperatures.
The barrier-lowering effect also allowed the ultimate theoretical strength
of the material to be included in a self-consistent way;
this was used in example simulations showing an abrupt increase in strain rate
for sufficiently high applied stress.

For the purposes of demonstration, we estimated the frequency factor
in the Arrhenius rate for dislocation motion from the Debye temperature,
which is based on a simplified representation of the vibrational mode structure
in the crystal lattice, and depends on compression but not temperature.
More rigorously, the frequency factor could be calculated from the 
population of phonon modes which is able to promote a particular hop of a
dislocation: a quantity which depends on temperature as well as compression.

Although constructed with dislocations in mind, the treatment of
defects is more general, and may be a suitable framework for deformation
by twinning.

Ta was used as an example to illustrate the general behavior of the model,
and to suggest ways to obtain parameters for specific materials using
more detailed theories such as electronic structure, and experimental
measurements of basic material properties.
The calibration of parameters for any material is an involved process:
detailed treatments of specific materials will be described separately.
There is also much scope for future studies using 
molecular dynamics and dislocation dynamics to investigate 
the evolution of dislocations and pinning centers.

\section{Acknowledgments}
We would like to thank Jaime Marian for valuable discussions about dislocations.
Sheng-Nian Luo made helpful suggestions regarding the content 
of this manuscript.
Funding was provided by the LANL Program Office for the
National Nuclear Security Administration's Campaign 10,
Inertial Confinement Fusion,
and by the LLNL Laboratory Directed Research and Development Program,
under project 08-ER-038.
The work was performed under the auspices of
the U.S. Department of Energy under contracts W-7405-ENG-36,
DE-AC52-06NA25396, and DE-AC52-07NA27344.
P.~Peralta acknowledges funding from the
Department of Energy,
National Nuclear Security Administration, under grant DE-FG52-06NA26169.



\begin{thebibliography}{10}
\bibitem{Rashid92}{M.M.~Rashid and S.~Nemat-Nasser,
   Comput. Meth. in App. Mech. and Eng. {\bf 94}, pp~201-228 (1992).}
\bibitem{Schoenfeld98}{S.E.~Schoenfeld, 
   Int. J.~Plasticity {\bf 14}, 9, pp~871-890 (1998).}
\bibitem{Clayton05}{J.D.~Clayton,
   Int. J.~Solids Structures {\bf 42}, pp~4613-4640 (2005).}
\bibitem{Zhang05}{K.S.~Zhang, M.S.~Wu, and R.~Feng,
   Int. J.~Plasticity {\bf 21}, pp~801-834 (2005).}
\bibitem{Vogler08}{T.J.~Vogler and J.D.~Clayton,
   J.~Mech. Phys. Solids {\bf 56}, pp~297-335 (2008).}
\bibitem{Steinberg-Guinan}{D.J.~Steinberg, S.G.~Cochran and M.W.~Guinan,
   J.~Appl. Phys. {\bf 51}, 1496 (1980).}
\bibitem{Steinberg-Lund}{D.J.~Steinberg and C.M.~Lund,
   J.~Appl. Phys. {\bf 65}, 1528 (1989).}
\bibitem{MTS1}{H.~Mecking and U.F.~Kocks,
   Acta metall. {\bf 29}, 1865-1875 (1981).}
\bibitem{MTS2}{P.S.~Follansbee and U.F.~Kocks,
   Acta metall. {\bf 36}, 81-93 (1988).}
\bibitem{PTW}{D.L.~Preston, D.L.~Tonks, and D.C.~Wallace,
   J.~Appl. Phys. {\bf 93}, 211 (2003).}
\bibitem{Becker04}{R.~Becker,
   Int.~J. Plasticity {\bf 20}, 11, 1983-2006 (2004).}
\bibitem{Swift_pop}{D.C.~Swift, T.E.~Tierney, S.-N.~Luo, D.L.~Paisley,
   G.A.~Kyrala, A.~Hauer, S.R.~Greenfield, A.C.~Koskelo, K.J.~McClellan,
   H.E.~Lorenzana, D.~Kalantar, B.A.~Remington, P.~Peralta, and E.~Loomis,
   Phys. Plasmas {\bf 12}, 056308 (2005).}
\bibitem{Swift_Be_07}{D.C.~Swift, T.E.~Tierney, S.-N.~Luo, R.N.~Mulford,
   G.A.~Kyrala, R.P.~Johnson, J.A.~Cobble, D.L.~Tubbs, and N.M.~Hoffman,
   {\it Dynamic plasticity of beryllium in the inertial fusion fuel capsule
   regime}, preprint {\tt arXiv:0711.3017}.}
\bibitem{Bringa}{E.M.~Bringa, A.~Caro, Y.~Wang, M.~Victoria, J.M.~McNaney,
   B.A.~Remington, R.F.~Smith, B.R.~Torralva, and H.~Van~Swygenhoven,
   Science {\bf 309}, 5742, 1838-1841 (2005).}
\bibitem{Benson92}{D.~Benson,
   Computer Methods in Appl. Mechanics and Eng. {\bf 99}, 235 (1992).}
\bibitem{Swift APL}{D.C.~Swift and G.J.~Ackland,
   Appl. Phys. Lett. {\bf 83}, 6, 1151-1153 (2003).}
\bibitem{Ariadne}{Software and documentation for the `Ariadne' material
   models package, version 7.0
   (Wessex Scientific and Technical Services, Perth, 2007).}
\bibitem{Swift_genscalar_07}{D.C.~Swift,
   {\it Numerical solution of shock and ramp loading relations for general
      material properties}, preprint {\tt arXiv:0704.0008}.}
\bibitem{Poirier99}{J.P.~Poirier and G.D.~Price,
   Phys. of the Earth and Planetary Interiors {\bf 110} pp~147-56 (1999).}
\bibitem{Ogden}{R.W.~Ogden, ``Non-Linear Elastic Deformations''
   (Dover, New York, 1997).}
\bibitem{OranBoris}{E.S.~Oran and J.P.~Boris,
   ``Numerical Simulation of Reactive Flow''
   (Elsevier, New York, 1987).}
\bibitem{Fotiu96}{P.A.~Fotiu and S.~Nemat-Nasser,
   Comput. Struct. {\bf 59}, 6, 1173-1184 (1996).}
\bibitem{Jones62}{O.E.~Jones, F.W.~Nelson, and W.B.~Benedick,
   J.~Appl. Phys. {\bf 33}, 3324 (1962).}
\bibitem{Taylor63}{J.W.~Taylor and M.W.~Rice,
   J.~Appl. Phys. {\bf 34}, 364 (1963).}
\bibitem{Stepanov85}{G.V.~Stepanov and V.V.~Kharchenko,
   Strength of Materials {\bf 18}, 1 (1985).}
\bibitem{Resseguier98}{T.~de~Resseguier and M.~Hallouin,
   J.~Appl. Phys. {\bf 84}, 1932 (1998).}
\bibitem{Hill}{R.~Hill, ``The Mathematical Theory of Plasticity''
   (Clarendon Press, Oxford, 1950).}
\bibitem{Kinslow70}{R.~Kinslow (Ed.),
   ``High-Velocity Impact Phenomena''
   Academic Press, New York, 1970).}
\bibitem{SESAME}{K.S.~Holian (Ed.), Los Alamos National Laboratory
   report LA-10160-MS (1984).}
\bibitem{Steinberg96}{D.J.~Steinberg, {\it Equation of State and Strength
   Properties of Selected Materials}, Lawrence Livermore National Laboratory
   report UCRL-MA-106439 change 1 (1996).}
\bibitem{Eremets96}{M.I.~Eremets, ``High Pressure Experimental Methods''
   (Oxford University Press, New York, 1996).}
\bibitem{Swift PRB 01}{D.C.~Swift, G.J.~Ackland, A.~Hauer, and G.A.~Kyrala,
   Phys. Rev. B {\bf 64}, 214107 (2001).}
\bibitem{Moriarty02}{J.A.~Moriarty, J.F.~Belak, R.E.~Rudd, P.~S\"oderlind,
   F.H.~Streitz, and L.H.~Yang,
   J.~Phys.: Cond. Matt. {\bf 14}, 2825-2857 (2002.)}
\bibitem{Featherston63}{F.H.~Featherston and J.R.~Neighbours,
   Phys. Rev. {\bf 130}, 4, 1324-1333 (1963).}
\bibitem{Kalidindi98}{S.R.~Kalidindi, 
   J.~Mech. Phys. Solids {\bf 46}, 267-290 (1998).}
\bibitem{Staroselsky98}{A.~Staroselsky and L.~Anand,
   J.~Mech. Phys. Solids {\bf 46}, 671 (1998).}
\bibitem{Karaman00}{I.~Karaman, H.~Sehitoglu, A.~Beaudoin, Y.~Chumlyakov, 
   H.~Maier, and C.~Tom\'e, 
   Acta Materialia {\bf 48}, 2031 (2000).}
\bibitem{Salem05}{A.~Salem, S.~Kalidindi, and S.~Semiatin, 
   Acta Materialia {\bf 53}, 3495 (2005).}
\bibitem{Barton07}{N.~Barton and H.-R.~Wenk,
   Model. Simul. Mater. Sci. Eng. {\bf 15}, 369-384 (2007).}
\bibitem{NematNasser98}{S.~Nemat-Naser, T.~Okinaka, and L.~Ni,
   J.~Mech. Phys. Solids {\bf 46}, 6, pp~1009-1038 (1998).}
\bibitem{Nichols71}{F.A.~Nichols,
   Mater. Sci. Eng. {\bf 8}, pp~108-120 (1971).}
\bibitem{Deo05}{C.S.~Deo, D.J.~Srolovitz, W.~Cai, and V.V.~Bulatov,
   J.~Mech. Phys. Solids {\bf 53}, 1223-1247 (2005).}
\bibitem{Farren25}{W.S.~Farren and G.I.~Taylor,
   Proc. Roy. Soc. {\bf A107}, 422 (1925).}
\bibitem{Taylor34}{G.I.~Taylor and H.~Quinney,
   Proc. Roy. Soc. {\bf A143}, 307 (1934).}
\bibitem{Qin92}{Q.~Qin and J.L.~Bassani,
   J.~Mech. Phys. Solids {\bf 44}, 4, pp~813-833 (1992).}
\bibitem{Peralta05}{P.~Peralta, D.~Swift, E.~Loomis, C.H.~Lim,
   and K.J.~McClellan,
  Metall. and Mat. Trans. A, {\bf 36}, 6, 1459-1469 (2005).}
\bibitem{Hirth}{J.P.~Hirth and J.~Lothe,
   ``Theory of Dislocations''
   (McGraw-Hill, New York, 1983).}
\bibitem{Johnson in Asay}{J.N.~Johnson in J.R.~Asay and M.~Shahinpoor (Eds),
   ``High-Pressure Shock Compression of Solids''
   (Springer-Verlag, New York, 1993).}
\bibitem{Elbaum60}{C.~Elbaum, 
   J.~Appl. Phys. {\bf 31}, 8, pp~1413-1415 (1960).}
\bibitem{Tang08}{M.~Tang (Lawrence Livermore National Laboratory),
   private communication.}
\bibitem{Tierney08}{T.E.~Tierney and D.C.~Swift, in preparation (2008).}
\bibitem{Quinney37}{H.~Quinney and G.I.~Taylor,
   Proc. Roy. Soc. {\bf A 163}, 913, pp~157-181 (1937).}
\end{thebibliography}
\end{document}